\documentclass[11pt]{article}

\usepackage[T1]{fontenc}
\usepackage{lmodern}
\usepackage{amsmath,amssymb,mathtools}
\usepackage{amsthm}
\usepackage{algorithm}
\usepackage[noend]{algpseudocode}
\usepackage{xparse}
\usepackage{xspace}
\usepackage{microtype}
\usepackage[a4paper,margin=1in]{geometry}
\usepackage[hidelinks]{hyperref}
\usepackage{orcidlink}
\usepackage[nameinlink,noabbrev]{cleveref}

\crefname{theorem}{Theorem}{Theorems}
\Crefname{theorem}{Theorem}{Theorems}
\crefname{lemma}{Lemma}{Lemmas}
\Crefname{lemma}{Lemma}{Lemmas}
\crefname{corollary}{Corollary}{Corollaries}
\Crefname{corollary}{Corollary}{Corollaries}
\crefname{definition}{Definition}{Definitions}
\Crefname{definition}{Definition}{Definitions}
\usepackage{xcolor}
\providecommand{\doi}[1]{\url{https://doi.org/#1}}

\newtheorem{theorem}{Theorem}[section]
\newtheorem{lemma}[theorem]{Lemma}
\newtheorem{corollary}[theorem]{Corollary}
\theoremstyle{definition}
\newtheorem{definition}[theorem]{Definition}

\DeclarePairedDelimiter\abs{\lvert}{\rvert}

\DeclarePairedDelimiter{\Paren}{(}{)}
\DeclarePairedDelimiter{\Bracket}{[}{]}
\DeclarePairedDelimiter{\Brace}{\{}{\}}
\NewDocumentCommand{\Oh}{s o m}{%
  \ensuremath{\mathcal{O}\IfBooleanTF{#1}{\Paren*{#3}}{%
    \IfNoValueTF{#2}{\Paren{#3}}{\Paren[#2]{#3}}}}}
\NewDocumentCommand{\Om}{s o m}{%
  \ensuremath{\Omega\IfBooleanTF{#1}{\Paren*{#3}}{%
    \IfNoValueTF{#2}{\Paren{#3}}{\Paren[#2]{#3}}}}}
\NewDocumentCommand{\Th}{s o m}{%
  \ensuremath{\Theta\IfBooleanTF{#1}{\Paren*{#3}}{%
    \IfNoValueTF{#2}{\Paren{#3}}{\Paren[#2]{#3}}}}}
\NewDocumentCommand{\oh}{s o m}{%
  \ensuremath{o\IfBooleanTF{#1}{\Paren*{#3}}{%
    \IfNoValueTF{#2}{\Paren{#3}}{\Paren[#2]{#3}}}}}
\NewDocumentCommand{\om}{s o m}{%
  \ensuremath{\omega\IfBooleanTF{#1}{\Paren*{#3}}{%
    \IfNoValueTF{#2}{\Paren{#3}}{\Paren[#2]{#3}}}}}
\NewDocumentCommand{\Prob}{s o m}{%
  \ensuremath{\Pr\IfBooleanTF{#1}{\Bracket*{#3}}{%
    \IfNoValueTF{#2}{\Bracket{#3}}{\Bracket[#2]{#3}}}}}
\NewDocumentCommand{\E}{s o m}{%
  \ensuremath{\mathbb{E}\IfBooleanTF{#1}{\Bracket*{#3}}{%
    \IfNoValueTF{#2}{\Bracket{#3}}{\Bracket[#2]{#3}}}}}
\NewDocumentCommand{\Max}{s o m m}{%
  \ensuremath{\max\IfBooleanTF{#1}{\Brace*{#3,\,#4}}{%
    \IfNoValueTF{#2}{\Brace{#3,\,#4}}{\Brace[#2]{#3,\,#4}}}}}
\NewDocumentCommand{\Min}{s o m m}{%
  \ensuremath{\min\IfBooleanTF{#1}{\Brace*{#3,\,#4}}{%
    \IfNoValueTF{#2}{\Brace{#3,\,#4}}{\Brace[#2]{#3,\,#4}}}}}
\NewDocumentCommand{\Exp}{s o m}{%
  \ensuremath{\exp\IfBooleanTF{#1}{\Paren*{#3}}{%
    \IfNoValueTF{#2}{\Paren{#3}}{\Paren[#2]{#3}}}}}
\NewDocumentCommand{\Log}{s o m}{%
  \ensuremath{\log\IfBooleanTF{#1}{\Paren*{#3}}{%
    \IfNoValueTF{#2}{\Paren{#3}}{\Paren[#2]{#3}}}}}
\NewDocumentCommand{\Var}{s o m}{%
  \ensuremath{\mathrm{Var}\IfBooleanTF{#1}{\Bracket*{#3}}{%
    \IfNoValueTF{#2}{\Bracket{#3}}{\Bracket[#2]{#3}}}}}
\NewDocumentCommand{\SetBuilder}{s o m m}{%
  \ensuremath{\IfBooleanTF{#1}{\Brace*{\,#3:#4\,}}{%
    \IfNoValueTF{#2}{\Brace{\,#3:#4\,}}{\Brace[#2]{\,#3:#4\,}}}}}

\renewcommand{\phi}{\ensuremath{\varphi}}
\newcommand{\cF}{\mathcal F}
\newcommand{\INV}{\ensuremath{\mathrm{Inv}}\xspace}
\newcommand{\ind}{\mathbf 1}

\title{Tight Bounds for Distance-Unbiased Sorting}
\author{%
Mats Bierwirth\,\orcidlink{0009-0006-6470-4455}\\
{\small Sorbonne Université, CNRS, LIP6, Paris, France}\\
{\small\texttt{mats.bierwirth@lip6.fr}}\\[1.2em]
Carola Doerr\,\orcidlink{0000-0002-4981-3227}\\
{\small Sorbonne Université, CNRS, LIP6, Paris, France}\\
{\small\texttt{carola.doerr@lip6.fr}}\\[1.2em]
Johannes Lengler\,\orcidlink{0000-0003-0004-7629}\\
{\small Department of Computer Science, ETH Zürich, Switzerland}\\
{\small\texttt{johannes.lengler@inf.ethz.ch}}
}
\date{}

\hypersetup{
  pdftitle={Tight Bounds for Distance-Unbiased Sorting},
  pdfauthor={Mats Bierwirth, Carola Doerr, Johannes Lengler},
  pdfkeywords={runtime analysis, randomized local search,
    permutation-based optimization, parameter control, randomized sorting}
}

\begin{document}
\maketitle
\begin{abstract}
\noindent We study elitist randomized local search for sorting permutations. In this model, 
each iteration considers a possible exchange of two positions. The entries are exchanged if and only if they are in the wrong order. We introduce \emph{distance-unbiased}
selection rules, which may choose an arbitrary distribution over
exchange distances but sample uniformly among all pairs at each chosen
distance. For static rules, we prove that the optimal worst-case
high-probability runtime is $\Theta(n(\log n)^2)$. This establishes the
asymptotic optimality of the harmonic sorter within this class and
partially resolves an open question of Olesker-Taylor. We then allow the
distance distribution to change either according to a predetermined
schedule or adaptively using the current permutation and the entire
history. Matching upper and lower bounds show that both classes have
optimal runtime
$\Theta(n(\log n)^2/\log\log n)$. Thus, scheduled adjustments of the distance distribution improve the
optimal static worst-case runtime by a factor of
$\Theta(\log\log n)$, while full-information adaptation provides no
further asymptotic improvement.
\end{abstract}
\medskip
\noindent\textbf{Keywords:}
Runtime analysis; randomized local search; permutation-based optimization;
parameter control; randomized sorting

\section{Introduction}
\label{sec:introduction}
Few algorithmic problems have been studied as thoroughly and are as
well understood as sorting%
~\cite{CormenLeisersonRivestStein2022,Knuth1998Sorting}. Over decades,
researchers have developed sorting algorithms for different
computational models and resource constraints, drawing on a wide range
of data structures and algorithmic ideas. Among these, a natural family
can be expressed as elitist local search on the space of
permutations: starting from an initial permutation, the algorithm
repeatedly selects two positions and exchanges their entries whenever
they form an inversion. Different selection rules
then recover, for example, Bubble Sort and Selection Sort. 

Naturally, sorting as local search has been studied primarily within
the evolutionary computation community, where simple randomized
selection rules give rise to algorithms such as Swap-RLS and
Exchange-RLS~\cite{BaumannRutterSudholt2026,ScharnowTinnefeldWegener2004}. Still, several modern randomized sorting algorithms developed outside
this community, including Guess-Sort, probabilistic Bubble Sort,
Annealing Sort, and the recent harmonic sorter, can also be viewed
within the same framework%
~\cite{BiedlEtAl2004,GasieniecSpirakisStachowiak2023,Goodrich2014,OleskerTaylor2025}.

We study a restricted class of selection rules that we call
\emph{distance-unbiased}. Such rules may allocate probability arbitrarily across exchange
distances but must distribute the mass assigned to each distance
uniformly among its pairs.
This restriction grants the rule complete control over the distance
distribution while preventing it from favoring one pair over another
at the same distance. To the best of our knowledge, this symmetry has not previously been
studied as a unified class, although many of the randomized algorithms
mentioned share it.

A central example of a distance-unbiased rule is the harmonic sorter of
Olesker-Taylor~\cite{OleskerTaylor2025}. In every iteration, it
samples the next pair independently from a fixed distribution in which
the probability of a pair is inversely proportional to its distance.
This rule sorts every permutation in $\Oh{n(\log n)^2}$ iterations with
high probability. Olesker-Taylor asks whether every fixed distribution
over pairs admits a worst-case permutation requiring
$\Om{n(\log n)^2}$ iterations with probability tending to one~\cite[Section~1.2]{OleskerTaylor2025}. We prove such a lower bound for all
distance-unbiased distributions,
thereby establishing the asymptotic optimality of the harmonic sorter
among static distance-unbiased rules and partially resolving
Olesker-Taylor's open question.

This static result leaves open whether the same distance distribution
should be used throughout the run. Since long exchanges can quickly
move entries that are far from their target positions, whereas short
exchanges become more useful once the remaining disorder is local, it
seems natural to shift probability mass toward shorter distances as
sorting progresses. This gives rise to a natural parameter-control
problem~\cite{DoerrDoerr2020ParameterControl,EibenHM99,EinarssonEtAl2019}: how much can be gained over the best static rule by changing the distance distribution over time, and must these choices depend on the current permutation to achieve this gain?

To tackle this question, we distinguish three nested classes of
distance-unbiased rules. The \emph{static} rules considered above fix
one distance distribution and use it throughout the run.
\emph{Scheduled} rules fix an entire sequence of distance distributions
before the run, whereas \emph{adaptive} rules may choose the next
distribution using the current permutation and the entire search history.
These classes separate the benefit of changing the distribution over
time from the potential additional benefit of an explicit use of the information gained during the search.

We resolve this question within the
distance-unbiased model by matching a scheduled upper bound with an
adaptive lower bound. Since scheduled rules form a subclass of
adaptive rules, the optimal worst-case high-probability runtime in both
classes is
\[
 \Th*{\frac{n(\log n)^2}{\log\log n}}.
\]
Thus, a predetermined schedule captures the entire asymptotic benefit
of changing the distance distribution: it improves over the best
static rule by a factor of $\Th{\log\log n}$, while complete information
about the current permutation and history provides no further
asymptotic improvement.

The remainder of the paper is structured as follows.
\Cref{sec:model} formalizes the algorithmic model and introduces the
probabilistic tools used throughout. \Cref{sec:static} proves the
static lower bound by embedding independent blind-search processes~\cite{DRWW2008} into disjoint blocks of one permutation. To withstand
adaptation, \Cref{sec:adaptive-lower} instead uses blocks at many
distance scales and shows that serving all scales quickly requires too much
cumulative probability mass.  \Cref{sec:upper} matches this lower bound
with a predetermined sequence of decreasing exchange radii, each of
which confines the remaining disorder to an increasingly local region. 
Finally, \Cref{sec:conclusion} interprets the results and discusses the
remaining open problems.

\section{Preliminaries}
\label{sec:model}
Throughout, let $\log$ denote the natural logarithm, and let \emph{with high
probability} mean with failure probability $n^{-\Omega(1)}$.
\subsection{The Sorting Problem}
Fix $n\ge2$, write $[n]=\{1,\ldots,n\}$, and let $S_n$ denote the set of permutations of $[n]$.  For a permutation $\pi \in S_n$, we measure its \emph{sortedness} by the number of inversions,
\[
 \INV(\pi)
 :=\abs*{\SetBuilder*{(i,j)\in[n]^2}{i<j,\ \pi(i)>\pi(j)}}.
\]
The identity is the unique permutation with no inversions and hence the unique optimum.

Although $\INV$ is a global objective, whether an exchange improves it is determined entirely by the two exchanged entries. Moreover, every non-identity permutation contains an inversion and therefore also admits an improving exchange.
\begin{lemma}[{\cite{ScharnowTinnefeldWegener2004}}]
\label[lemma]{lem:acceptance}
Exchanging positions $i<j$ decreases $\INV$ if and only if $  \pi(i)>\pi(j). $
\end{lemma}

\begin{proof}
Put $a=\pi(i)$ and $b=\pi(j)$. If $a>b$, the exchange removes their mutual inversion. For each $i<k<j$, it also removes two further inversions when $b<\pi(k)<a$ and preserves the total involving $\pi(k)$ otherwise. Lastly, entries outside $[i,j]$ contribute the same total as before. Hence $\INV$ strictly decreases. If $a<b$, apply the same argument to the reverse exchange.
\end{proof}
\noindent The \emph{displacement} of a value is defined as the distance between its value and its position. Similarly, the \emph{displacement} of a permutation is defined as the maximum displacement over all values:
\[
 \Delta(\pi):=\max_{i\in[n]}\abs{\pi(i)-i}.
\]
An exchange decreasing $\INV$ cannot increase the maximum displacement.
\begin{lemma}
\label[lemma]{lem:displacement}
If an exchange decreasing $\INV$ is accepted, then it cannot increase~$\Delta$.
\end{lemma}

\begin{proof}
Suppose that entries $a>b$ at positions $i<j$ are exchanged, and put
\[
 D=\max\{\abs{a-i},\abs{b-j}\}.
\]
Then
\[
 i-D\le j-D\le b<a\le i+D\le j+D.
\]
Consequently, $\abs{b-i}\le D$ and $\abs{a-j}\le D$. The displacements of all other entries are unchanged.
\end{proof}

\subsection{The Generic-Exchange-RLS}

We study a class of elitist algorithms that modify a permutation by one exchange at a time, starting from an initial permutation $\pi_0$. Every algorithm in the class follows the same update rule and is determined by a possibly parameterized randomized \emph{sampling rule} $\mathcal S$.
\begin{algorithm}[H]
\caption{\textsc{Generic-Exchange-RLS}($\mathcal S$)}
\label{alg:exchange}
\begin{algorithmic}[1]
\For{$t=0,1,\ldots$}
    \State $(i,j)\gets\mathcal S$, where $i<j$
    \State $\displaystyle
    \pi_{t+1}\gets
    \begin{cases}
        \pi_t\circ(i,j), & \text{if }\pi_t(i)>\pi_t(j),\\
        \pi_t,            & \text{otherwise}
    \end{cases}$
\EndFor
\end{algorithmic}
\end{algorithm}
\noindent We call a sampling rule \emph{distance-unbiased} if, conditional on the complete process history, it assigns the same probability to all pairs of the same distance. Equivalently, drawing from $\mathcal S$ can be modeled as a two-step process: first sample a distance $d$ from a possibly parameterized distribution $p_t$ on $[n-1]$, then select the pair $(i,i+d)$ uniformly among all pairs of distance $d$.

For every $d\in[n-1]$ and $i\in[n-d]$, the conditional probability of sampling the pair $(i,i+d)$ in the next iteration is therefore given by
\begin{equation}
 \Prob*{\mathcal S\text{ returns }(i,i+d)\mid\cF_t}
 =\frac{p_t(d)}{n-d},
 \label{eq:kernel}
\end{equation}
where $\cF_t$ contains the current permutation and the process history through iteration $t$.

We distinguish three nested classes. In the first two, distributions may depend on $n$ and the algorithm's parameters, but not on the initialization $\pi_0$.
\begin{description}
 \item[Static.]
 The same distribution $p$ is used in every iteration.

 \item[Scheduled.]
 The distribution sequence $(p_t)_{t\ge0}$ is fixed before the run.

 \item[Adaptive.]
 The choice of $p_t$ may depend on the current permutation and the  entire process history.
\end{description}
Many natural sampling rules such as Swap-RLS and Exchange-RLS~\cite{BaumannRutterSudholt2026,ScharnowTinnefeldWegener2004}, uniform bounded-radius exchange~\cite{Cicirello2014}, and the recent harmonic sorter by Olesker-Taylor~\cite{OleskerTaylor2025} fall into the static class.

\subsection{Runtime}

For a Generic-Exchange-RLS with sampling rule $\mathcal S$, initialized with $\pi_0=\pi$, define
\[
 T_{\mathcal S}(\pi):=\inf\{t\ge0:\pi_t=\mathrm{id}\}
\]
as its \emph{runtime}, measured in proposed exchanges, including rejected ones. If the identity is never reached, we set $T_{\mathcal S}(\pi)=\infty$. For a static distance law $p$, we also write $T_p(\pi)$.

For convenience, we let the process continue after it reaches the identity. Since the identity contains no inversions, every subsequent 
proposed exchange leaves the permutation unchanged.

\subsection{Two Counting Bounds}

We use two standard counting bounds. They follow from binomial comparison and Chernoff bounds~\cite[Lemma~1.8.7 and Eqs.~(1.10.7) and~(1.10.12)]{Doerr2020Probabilistic}.

\begin{lemma}
\label[lemma]{lem:counting}
Let $t,k\ge1$ be integers, $q\in[0,1]$, and $Z\sim\operatorname{Bin}(t,q)$. Let $Y_i$ indicate success in iteration $i$, and write $\cF_i$ for the information available before sampling in iteration $i+1$.
\begin{enumerate}
 \renewcommand{\labelenumi}{\textup{(\roman{enumi})}}
 \item Suppose $\Prob{Y_i=1\mid\cF_{i-1}}\le q$ for all  $i\in[t]$. If $k\ge tq$, then
 \[
  \Prob*{\sum_{i=1}^tY_i\ge k}
  \le\Prob{Z\ge k}
  \le\left(\frac{etq}{k}\right)^k.
 \]

 \item Suppose $\Prob{Y_i=1\mid\cF_{i-1}}\ge q$ for all  $i\in[t]$. If $k\le tq/2$, then
 \[
  \Prob*{\sum_{i=1}^tY_i<k}
  \le\Prob{Z<k}
  \le e^{-tq/8}.
 \]
\end{enumerate}
\end{lemma}

\section{A Lower Bound for Static Rules}
\label{sec:static}
In this section we prove an $\Om{n(\log n)^2}$ runtime lower bound for static distance-unbiased sampling rules by reducing sorting to blind search on the integers. The bound is asymptotically tight by \Cref{cor:static-optimal}.
\begin{definition}[Blind Search]
Fix $m\ge2$ and a probability distribution $\mu$ on $[m]$. The \textit{blind-search process} starts from $X_0\sim\operatorname{Unif}([m])$. In each step it independently samples $d\sim\mu$ and replaces the current state $X$ by $X-d$ if $d\le X$; otherwise the state remains unchanged. Its hitting time is
\[
 T_{\mathrm{BS}}:=\inf\{t\ge0:X_t=0\}.
\]
\end{definition}

\noindent The lower-bound argument for the expected runtime of this process by Dietzfelbinger et al.~\cite[Section~3.3]{DRWW2008} can be strengthened to a constant-probability form.

\begin{lemma}
\label[lemma]{lem:blind-search}
There is an absolute constant $c_{\mathrm{BS}}>0$ such that, for every $m\ge2$ and every distribution $\mu$ on $[m]$ with $\mu(1)>0$,
\[
 \Prob{T_{\mathrm{BS}}>c_{\mathrm{BS}}(\log m)^2}\ge\frac12.
\]
\end{lemma}
\begin{proof}
The deferred-decision reformulation of Dietzfelbinger et al.~\cite[Section~3.2]{DRWW2008} gives a step-by-step coupling of blind search on the integers with a chain $S$. Both processes use the same proposals. Conditional on $S=s>0$, the blind-search position is uniform on $[s]$, and $S=0$ exactly when the position is $0$. Both processes are therefore absorbed in the same step.

For this chain, the potential of Dietzfelbinger et al.~\cite[Definition~3.3 and Lemmas~3.4--3.5]{DRWW2008} is nonnegative and nonincreasing, vanishes at absorption, and satisfies
\[
 \Phi(m)\ge c_0(\log m)^2,
 \qquad
 \E{\Phi(S_i)-\Phi(S_{i+1})\mid S_i}\le C_0
\]
for absolute constants $c_0,C_0>0$. Summing the one-step drift bounds and applying Markov's inequality gives
\[
 c_0(\log m)^2\Prob{T_{\mathrm{BS}}\le k}
 \le \E{\Phi(m)-\Phi(S_k)}
 \le C_0k.
\]
Taking $k=\lfloor c_0(\log m)^2/(2C_0)\rfloor$ proves the claim with $c_{\mathrm{BS}}=c_0/(2C_0)$. We refer to~\cite[Sections~3.2--3.4]{DRWW2008} for a more detailed explanation of the chain construction as well as the potential estimate.
\end{proof}

\noindent We embed $K=\lceil\log n\rceil$ blind-search processes into disjoint blocks of one permutation. Sorting requires every process to finish, whereas \Cref{lem:blind-search} shows that each process has constant probability of requiring many activations. Consequently, the probability that all blocks finish early is polynomially small.

\begin{theorem}
\label[theorem]{thm:static-lower}
There is an absolute constant $c>0$ such that, for every sufficiently large $n$ and for every static distance-unbiased sampling rule with distance distribution $p$, there exists a permutation $\pi\in S_n$ satisfying
\[
 \Prob{T_p(\pi)>c n(\log n)^2}
 \ge 1-n^{-c}.
\]
\end{theorem}
\begin{proof}
Fix $p$. If $p(1)=0$, the permutation $(2,1,3,\ldots,n)$ never changes, since its only inversion has distance $1$. The theorem is then immediate. Hence assume $p(1)>0$. Set
\[
 K=\lceil\log n\rceil,
 \qquad
 m=\left\lfloor\frac{n}{K+1}\right\rfloor-1.
\]
Then $K=\Th{\log n}$ and $m=\Th{n/\log n}$. Begin with an identity prefix of length $m$, followed by $K$ consecutive \emph{blocks} of length $m+1$, leaving all remaining positions fixed. For each block occupying positions $a,\ldots,a+m$, independently choose $X_j\sim\operatorname{Unif}([m])$ and place its values in the order
\[
 (a+1,\ldots,a+X_j,a,a+X_j+1,\ldots,a+m).
\]
Thus the block minimum $a$ has initial displacement $X_j$, matching the uniform initial state in the blind-search process.

Every value in an earlier block is smaller than every value in a later block. Hence every cross-block proposal is rejected. Since exchanges within a block preserve its set of values, this remains true throughout the run. Consequently, blocks do not interfere.

We track the minimum of each block. If its current displacement is $x$, then a proposal having this minimum as its right endpoint moves it left by $d$ exactly when $d\le x$. For $d>x$, the left endpoint lies before the block and contains a smaller value, so the proposal is rejected. The initial prefix has length $m$, so all distances in $[m]$ are available even for the first block.

Call a proposal of length at most $m$ whose right endpoint is a block minimum an \emph{activation}. In every iteration, conditional on the history, a fixed block is activated with probability
\[
 \lambda=\sum_{d=1}^m\frac{p(d)}{n-d}
 \le\frac1{n-m}\le\frac2n,
\]
for sufficiently large $n$. Conditional on an activation, its distance has distribution
\[
 \mu(d)=\frac{p(d)}{(n-d)\lambda}.
\]
Since $p(1)>0$, we have $\lambda>0$ and $\mu(1)>0$. Measured in activations, each minimum follows blind search with proposal law $\mu$. We may equivalently pre-sample an independent $\mu$-distributed proposal sequence for each block and reveal its next proposal whenever that block is activated.

Put
\[
 k=\lfloor c_{\mathrm{BS}}(\log m)^2\rfloor,
 \qquad
 t=\lfloor nk/8\rfloor.
\]
For each block $j$, let $B_j$ be the event that its minimum has not reached its target after its first $k$ activations. The initial displacements and proposal sequences are independent across blocks. Thus, by \Cref{lem:blind-search}, the events $B_1,\ldots,B_K$ are independent and satisfy
\[
 \Prob{B_j}\ge\frac12.
\]
Let $N_j$ denote the number of times block $j$ is activated during the first $t$ iterations. For each fixed $j$,
\[
 N_j\sim\operatorname{Bin}(t,\lambda).
\]
Since $t\lambda\le k/4$, \Cref{lem:counting}(i) gives
\[
 \Prob{N_j>k}\le\left(\frac e4\right)^k.
\]
If the permutation is sorted by time $t$, then either no event $B_j$ occurs or some block has received more than $k$ activations. Thus,
\[
 \Prob{T_p(\pi)\le t}
 \le \Prob{\text{no }B_j\text{ occurs}}
       +\sum_{j=1}^K\Prob{N_j>k}
 \le 2^{-K}+K(e/4)^k.
\]
By choice of $K$, $k$, and $t$, the right-hand side is $n^{-\Theta(1)}$. By the probabilistic method, some deterministic choice of $X_1,\ldots,X_K$ satisfies the same bound. Choosing $c>0$ sufficiently small proves the theorem.
\end{proof}

\noindent Combining \Cref{thm:static-lower} with the matching  bound for the harmonic sorter of Olesker-Taylor~\cite[Theorem~1.5]{OleskerTaylor2025} gives the following tight bound.

\begin{corollary}
\label[corollary]{cor:static-optimal}
For static distance-unbiased rules, the optimal worst-case high-probability runtime for the sorting problem is $\Th*{n(\log n)^2}$.
\end{corollary}

\section{A Lower Bound for Adaptive Rules}
\label{sec:adaptive-lower}
The static blind-search argument does not extend to scheduled rules. Proposing the powers of two up to $m$ in decreasing order reaches zero from every starting state in $[m]$ in $\Oh{\log m}$ proposals. The optimal expected runtime under a fixed law is $\Th{(\log m)^2}$~\cite{DRWW2008}. This logarithmic speedup, however, does not transfer to the sorting problem. In fact, a modified version of the same lower bound construction from the previous section gives an adaptive lower bound of $\Om*{\frac{n(\log n)^2}{\log\log n}}$, only a factor of $\Th{\log\log n}$ below the static lower bound.

As in the static construction, we divide the permutation into blocks and track their minima. This time, we use blocks of different sizes.

\begin{definition}[Block Construction]
\label[definition]{def:adaptive-blocks}
For sufficiently large $n$, set
\[
 G=\lceil(\log n)^3\rceil,\qquad
 h=\left\lfloor\frac{\log n}{3\log G}\right\rfloor,\qquad
 K=\lceil\sqrt n\rceil,
\]
and let $m_s=G^s$ for $s=0,\ldots,h$. Each scale $s\in[h]$ consists of $K$ blocks of length $m_s+1$.  To construct $\pi$, place the blocks consecutively, starting at position $1$ and taking the scales in increasing order. For each block occupying positions $a,\ldots,a+m_s$, place its values in the order
\[(a+1,\ldots,a+m_s,a).\]
Fill each remaining position with its own value and treat that position as a singleton block.
\end{definition}
\noindent We have $m_h\le n^{1/3}$ and $h=\Th{\log n/\log\log n}$, so the blocks at all scales occupy at most $Kh(m_h+1)=\oh{n}$ positions and $\pi$ is well defined. As before, every value in an earlier block is smaller than every value in a later block. Hence every cross-block proposal is rejected. Since exchanges within a block preserve its set of values, this remains true throughout the run. We track only the minima of the $K$ blocks at each scale.

Assign each scale a \emph{distance range} 
\[
 \mathcal D_s=\{m_{s-1}+1,\ldots,m_s\}.
\]
A minimum at scale $s$ starts $m_s$ positions to the right of its target and can only move left within its block. Proposals longer than $m_s$ cannot move it. If it never moves by a distance in $\mathcal D_s$, each move covers at most $m_{s-1}$ positions, so it takes at least $G=m_s/m_{s-1}$ moves to reach its target. The first lemma bounds the chance that any tracked minimum makes that many moves by the deadline. On the event that no block does, sorting requires every block minimum to make at least one move in its own distance range. In a second lemma we show that, with high probability, each scale achieves this only after sufficient cumulative probability mass has been assigned to that scale. We finish the argument by showing that below the proposed runtime bound, not enough mass exists to satisfy all scales.

\begin{lemma}
\label[lemma]{lem:adaptive-move-count}
For any adaptive distance-unbiased sampling rule started from $\pi$ and every fixed integer $1\le t\le nh\log K/4$, the probability that some block minimum makes at least $G$ moves by time $t$ is at most $n^{-1}$.
\end{lemma}
\begin{proof}
A block minimum can only move left, so at distance $d$ at most one pair can move it. Every such move has $d\le m_h\le n/2$, and by \eqref{eq:kernel} that pair is proposed with conditional probability $p_i(d)/(n-d)\le 2p_i(d)/n$ in iteration $i+1$. Summing over distances, a fixed minimum moves with conditional probability at most $2/n$ per iteration.

Since $t/n=\Oh{(\log n)^2}$ and $G\ge(\log n)^3$, we have $2et/(nG)\le 1/2$ for sufficiently large $n$. By \Cref{lem:counting}(i) and a union bound over the block minima,
\[
 \Prob{\text{some block minimum makes at least $G$ moves by time $t$}}
 \le n\left(\frac{2et}{nG}\right)^G.
\]
This is at most $n\,2^{-G}\le n^{-1}$ for large $n$.
\end{proof}

\noindent If every block minimum makes fewer than $G$ moves, sorting requires each one to move in its own distance range. Let $U_s(t)$ count the minima at scale $s$ still waiting for their first move in $\mathcal D_s$ after $t$ iterations.

\begin{lemma}
\label[lemma]{lem:exposure}
For any adaptive distance-unbiased sampling rule started from $\pi$ and every fixed integer $1\le t\le nh\log K/4$,
\[
 \Prob{U_s(t)=0\text{ for all }s\in[h]}\le hK^{-1/2}.
\]
\end{lemma}
\begin{proof}
For each scale $s$, write
\[
 \Lambda_s(t)=\sum_{i=0}^{t-1}\sum_{d\in\mathcal D_s}p_i(d)
\]
for the \emph{cumulative probability mass} assigned to $\mathcal D_s$ in the first $t$ iterations. The ranges are disjoint, so $\sum_{s=1}^h\Lambda_s(t)\le t$. If $U_s(t)=0$ for every scale, then some scale has both $U_s(t)=0$ and $\Lambda_s(t)\le t/h\le n\log K/4$. We bound this joint event for each fixed scale and then take a union bound.

Fix a scale $s$. In iteration $i+1$, let $q=\Lambda_s(i+1)-\Lambda_s(i)$ be the mass assigned to its range. By the pair estimate in the proof of \Cref{lem:adaptive-move-count}, each waiting minimum moves in its range with conditional probability at most $2q/n$. At most one minimum can move per exchange, so the count drops by one with conditional probability at most $2qU_s(i)/n$.

To bound the probability of finishing on little mass, consider the potential
\[
 M_i=\frac{e^{-2\Lambda_s(i)/n}}{1+U_s(i)}.
\]
If the count stays fixed, which it always does when $U_s(i)=0$, then $M_{i+1}=e^{-2q/n}M_i\le M_i$. If $U_s(i)>0$, a drop adds $e^{-2q/n}M_i/U_s(i)$ to this value. Since $q$ is known given $\cF_i$, the drop probability bound then gives
\begin{align*}
 \E{M_{i+1}\mid\cF_i}
 &\le e^{-2q/n}M_i
      +\frac{2qU_s(i)}{n}\cdot\frac{e^{-2q/n}M_i}{U_s(i)}\\
 &=e^{-2q/n}(1+2q/n)M_i\le M_i.
\end{align*}
Thus the potential has nonpositive conditional drift and $\E{M_t}\le M_0=1/(K+1)$.

If $U_s(t)=0$ and $\Lambda_s(t)\le n\log K/4$, then $M_t\ge K^{-1/2}$. Markov's inequality therefore bounds the probability of this event by $\sqrt K\,\E{M_t}\le K^{-1/2}$. A union bound over the $h$ scales proves the claim.
\end{proof}

\noindent Combining the two bounds gives the claimed high-probability runtime lower bound.

\begin{theorem}
\label[theorem]{thm:adaptive-lower}
There is an absolute constant $c>0$ such that, for every sufficiently large $n$ and every adaptive distance-unbiased sampling rule $\mathcal S$, the permutation $\pi$ from \Cref{def:adaptive-blocks} satisfies
\[
 \Prob*{T_{\mathcal S}(\pi)>
        c\,\frac{n(\log n)^2}{\log\log n}}
 \ge 1-n^{-c}.
\]
\end{theorem}
\begin{proof}
Fix an adaptive distance-unbiased sampling rule $\mathcal S$ and put
\[
 t=\left\lfloor\frac{nh\log K}{4}\right\rfloor.
\]
If sorting finishes by time $t$ and every block minimum has made fewer than $G$ moves, then each has moved in its own distance range, so $U_s(t)=0$ for all $s\in[h]$. By \Cref{lem:adaptive-move-count,lem:exposure} and a union bound,
\[
 \Prob{T_{\mathcal S}(\pi)\le t}\le n^{-1}+hK^{-1/2}.
\]
By choice of parameters, $t=\Th*{n(\log n)^2/\log\log n}$ and the right-hand side is $\Oh{n^{-1/4}\log n}$. Choosing $c>0$ sufficiently small gives the claimed bound.
\end{proof}

\section{A Matching Scheduled Rule}
\label{sec:upper}

We now construct a scheduled sampling rule for the sorting problem and prove its worst-case high-probability runtime bound.

For $r\in[n-1]$, Cicirello's window-limited swap operator~\cite[Section~3.1]{Cicirello2014} samples uniformly among all pairs at distance at most $r$. We call this the \emph{$r$-rule}. Its distance distribution is
\[
 p^{[r]}(d)=\frac{n-d}{Z_r}\ind_{\{d\le r\}},
 \qquad Z_r=\sum_{d=1}^r(n-d).
\]
Each eligible pair is proposed with probability $1/Z_r$.

We build our scheduled rule from a predefined sequence of $r$-rules with geometrically decreasing radii, inspired by the annealing strategy of Goodrich's Annealing Sort~\cite{Goodrich2014}.
\begin{definition}
\label[definition]{def:scheduled-rule}
Fix integers $R\ge2$, $h\ge0$, and $\tau\ge1$. The sampling rule $\mathcal S(R,h,\tau)$ uses $h$ phases. In phase $i=1,\ldots,h$, it uses the $r_i$-rule for $\tau$ iterations, where
\[
 r_i=\min\left\{
 2\left\lceil\frac{n}{R^i}\right\rceil,n-1
 \right\}.
\]
After these phases, it samples uniformly from all adjacent exchanges. As all parameters are fixed before the run, this rule is in the scheduled class.
\end{definition}

\noindent To choose the parameters correctly, we first analyze how one phase reduces displacement. We do this through binary threshold projections.

\begin{definition}[Threshold Projection~{\cite{OleskerTaylor2025}}]
\label[definition]{def:threshold-projection}
For $\pi\in S_n$ and $c\in[n-1]$, the \textit{threshold projection} of $\pi$ at $c$ replaces values at most $c$ by $0$ and larger values by $1$. Its sorted form has zeros in positions $1,\ldots,c$ and ones afterward.
\end{definition}

\noindent By the zero-one principle~\cite[Theorem~2.1 and Lemma~3.1]{Goodrich2011}, a permutation is sorted if all its threshold projections are sorted. This fact lets us analyze each projection separately and bound the total failure probability by a union bound. 

The advantages of threshold projections become apparent when considering their intimate relation to displacement: a displacement bound confines all misplaced bits in each projection to a window around its threshold $c$. Conversely, controlling these windows across all thresholds controls the maximum displacement. To study these windows uniformly, even near the ends of the array, we pad each projection by fixed zeros at $i\le0$ and fixed ones at $i>n$. These additional positions are never sampled, so padding does not change the algorithm. Furthermore, the padded entries are already correctly ordered, so they cannot participate in an improving exchange. 

We now formalize the relation between displacement and threshold windows.

\begin{lemma}
\label[lemma]{lem:localisation}
For any $\pi\in S_n$ and integer $D\ge0$, $\Delta(\pi)\le D$ if and only if, for every threshold $c\in[n-1]$, all misplaced bits lie in $[c-D+1,c+D]$. Each such window then contains exactly $D$ zeros and $D$ ones.
\end{lemma}
\begin{proof}
If $\Delta(\pi)\le D$, then $i\le c-D$ implies $\pi(i)\le i+D\le c$, while $i\ge c+D+1$ implies $\pi(i)\ge i-D>c$. Conversely, if $\pi(i)-i>D$, the threshold $c=\pi(i)-1$ has a misplaced one at $i\le c-D$. If $i-\pi(i)>D$, the threshold $c=\pi(i)$ has a misplaced zero at $i\ge c+D+1$.

Finally, the projection and its sorted form have the same number of zeros in the original array. Since they agree outside the window and the sorted window is perfectly balanced, the projection's window contains $D$ zeros and $D$ ones.
\end{proof}

\noindent We analyze one phase through its threshold projections. For each threshold, we first show that the rule reduces the total number of misplaced bits in its window. Once sufficiently few remain, the same phase moves them into a smaller window around the threshold. When this holds for all thresholds simultaneously, by \Cref{lem:localisation}, the maximum displacement must also decrease.

\begin{lemma}
\label[lemma]{lem:contraction}
There is an absolute constant $C>0$ such that the following holds. Let $R\ge2$, $R^2\le b<n$, and $t'\ge Cn(R^4+\log n)$ be integers, and suppose the current permutation $\pi_t$ satisfies $\Delta(\pi_t)\le Rb$. If the next $t'$ iterations use the $r$-rule with $r=\min\{2b,n-1\}$, then
\[
 \Prob{\Delta(\pi_{t+t'})>b\mid\cF_t}\le n^{-2}.
\]
\end{lemma}

\begin{proof}
Condition on $\cF_t$ and fix a threshold $c$. By \Cref{lem:displacement,lem:localisation}, all misplaced bits remain in the balanced padded window $[c-Rb+1,c+Rb]$. Divide this window into $2R$ blocks of size $b$ each, numbered from left to right, with the sorted boundary between the two central blocks. The projection changes only when a pair $10$ is proposed, in which case it becomes $01$. Every such pair between neighboring blocks has distance at most $2b-1$, so it is eligible under the $r$-rule and is proposed with probability $1/Z_r\ge1/(2nb)$.

We first show that after $\Oh{n(R^4+\log n)}$ iterations at most $b$ misplaced bits are left with high probability. For each time $s$ during the phase, let $u_i(s)$ count the ones in block $i$, and define
\[
 U(s)=\sum_{i=1}^{R}u_i(s)
     =\sum_{i=R+1}^{2R}(b-u_i(s)),
\]
as the total number of ones to the left of the threshold. Since the window is balanced, $2U(s)$ then counts all misplaced bits. Since accepted proposals never increase the number of misplaced bits, $U(s)$ cannot increase with time.

While $U(s)>b/2$, averaging gives a left block with more than $b/(2R)$ ones and a right block with more than $b/(2R)$ zeros. Since any block with at most $b/(2R)$ zeros has at least $b/(2R)$ ones, an intermediate-value argument gives neighboring blocks $i,i+1$ satisfying
\[
 u_i(s)(b-u_{i+1}(s))\ge\frac{b^2}{4R^2}.
\]
Call an exchange of a pair $10$ between neighboring blocks a \emph{success}. While $U(s)>b/2$, the preceding bounds give a conditional success probability of at least 
\[
 q=\frac{b}{8nR^2}.
\]
A success moves a one into the next block even if $U(s)$ stays unchanged. We measure this progress by
\[
 V(s)=\sum_{i=1}^{2R}i\,u_i(s).
\]
Each success increases $V(s)$ by one, and no accepted exchange decreases it. Since the window contains $Rb$ ones, we have $V(s)\le2R^2b$. Thus within $2R^2b$ successes, $U(s)$ must fall to at most $b/2$. Set
\[
 t_1=t+\left\lceil C_1n(R^4+\log n)\right\rceil.
\]
Then by \Cref{lem:counting}(ii), $U(s) \le b/2$ is reached by $t_1$ except with probability
\[
 \Prob{U(t_1)>b/2\mid\cF_t}\le\tfrac12n^{-3}
\]
for a sufficiently large absolute constant $C_1$.

We now show that $\Oh{n(R+\log n)}$ further iterations leave no misplaced bits outside the central two blocks. Condition on $\cF_{t_1}$ and assume $U(t_1)\le b/2$. Every left block then contains at least $b/2$ zeros and every right block at least $b/2$ ones. These bounds persist for the remaining argument because $U(s)$ cannot increase. We measure the remaining misplacement outside the central two blocks by
\[
 W(s)=\sum_{i=1}^{R-1}2^{R-i}u_i(s)
     +\sum_{i=R+2}^{2R}2^{i-R-1}(b-u_i(s)).
\]
Thus $W(s)=0$ exactly when all misplaced bits lie in the central two blocks. Accepted exchanges cannot increase $W(s)$.

Each misplaced one counted in the first sum has at least $b/2$ zero partners in the next block. Exchanging it with one of them reduces its contribution to $W(s)$ by at least half. The same holds for misplaced zeros in the second sum, using partners containing ones in the preceding block. Each pair is proposed with probability at least $1/(2nb)$, and no pair is counted twice. Hence
\[
 \E{W(s)-W(s+1)\mid\cF_s}\ge\frac{W(s)}{8n}.
\]
At most $b$ misplaced bits remain at time $t_1$, so $W(t_1)\le n2^R$. By the multiplicative-drift theorem with tail bounds~\cite[Theorem~2.4.5]{Lengler2020Drift}, $\Oh{n(R+\log n)}$ further iterations suffice to make $W$ zero, except with probability at most $\tfrac12n^{-3}$.

For sufficiently large $C$, both parts fit within the next $t'$ iterations. Combining their failure probabilities, all misplaced bits in each projection lie in its central two blocks at time $t+t'$ except with conditional probability at most $n^{-3}$ given $\cF_t$. A union bound over the $n-1$ thresholds and \Cref{lem:localisation} prove the claim.
\end{proof}

\noindent We now choose the parameters and analyze the sampling rule from \Cref{def:scheduled-rule}.

\begin{theorem}
\label[theorem]{thm:scheduled-upper}
There are parameter choices $R,h,\tau$ depending only on $n$ and an absolute constant $C>0$ such that, for every sufficiently large $n$ and every $\pi\in S_n$, the rule $\mathcal S=\mathcal S(R,h,\tau)$ satisfies
\[
 \Prob*{T_{\mathcal S}(\pi)\le
 C\frac{n(\log n)^2}{\log\log n}}
 \ge 1-\frac1n.
\]
\end{theorem}
\begin{proof}
For sufficiently large $n$, choose
\[
 R=\lceil(\log n)^{1/4}\rceil,
 \qquad \tau=\lceil C_0n\log n\rceil,
\]
where $C_0>0$ is a sufficiently large absolute constant. Put $D_i=\lceil n/R^i\rceil$ for $i\ge0$, and choose $h$ as the first index with $D_h\le R^4$. We show that phase $i$ reduces the displacement bound from $D_{i-1}$ to $D_i$.

Suppose phase $i\in[h]$ starts with displacement at most $D_{i-1}$, and put $b=D_i$. The phase uses radius $\min\{2b,n-1\}$. Since $R$ is an integer, $D_{i-1}\le Rb$. By the choice of $h$, we have $D_{i-1}>R^4$, and hence $b>R^3\ge R^2$. By \Cref{lem:contraction}, the phase ends with displacement at most $D_i$ except with probability $n^{-2}$. Since $R^4=\Th{\log n}$, choosing $C_0$ large enough makes $\tau$ sufficient for every contraction phase.

Initially $\Delta(\pi_0)\le n=D_0$. If all contraction phases succeed, the final adjacent-exchange phase starts with displacement at most $D_h\le R^4$. Every inversion $i<j$ at that time satisfies $j-R^4\le\pi_{h\tau}(j)<\pi_{h\tau}(i)\le i+R^4$, so $j-i<2R^4$. This phase is Swap-RLS, so Theorem~6.11 of Baumann et al.~\cite{BaumannRutterSudholt2026} gives a conditional failure probability at most $n^{-2}$ within $\Oh{n(R^4+\log n)}=\Oh{n\log n}$ iterations. One sufficiently large choice of $C_0$ therefore covers all phases.

The number of contraction phases and the total time bound satisfy
\[
 h=\Oh*{\frac{\log n}{\log\log n}},
 \qquad
 (h+1)\tau=\Oh*{\frac{n(\log n)^2}{\log\log n}}.
\]
A union bound over the first failed phase gives uniformly over $\pi$
\[
 \Prob{T_{\mathcal S}(\pi)>(h+1)\tau}
 \le (h+1)n^{-2}\le\frac1n
\]
for sufficiently large $n$. Choosing $C$ sufficiently large proves the claim.
\end{proof}

\noindent Since scheduled rules form a subclass of adaptive rules, \Cref{thm:adaptive-lower,thm:scheduled-upper} give the following tight bound.

\begin{corollary}
\label[corollary]{cor:time-varying-adaptive-optimal}
For both scheduled and adaptive distance-unbiased rules, the optimal worst-case high-probability runtime for the sorting problem is $\Th*{\frac{n(\log n)^2}{\log\log n}}$.
\end{corollary}

\section{Conclusion}
\label{sec:conclusion}

We determine the optimal worst-case high-probability runtime of static
distance-unbiased sorting rules as $\Th{n(\log n)^2}$, establishing the
harmonic sorter's asymptotic optimality within this class and partially
resolving the open question of Olesker-Taylor~\cite{OleskerTaylor2025}.
For both scheduled and adaptive rules, the optimal worst-case high-probability runtime is
$\Th*{n(\log n)^2/\log\log n}$. Thus, a predetermined schedule yields
the entire factor-$\Th{\log\log n}$ improvement, while full-information
adaptation provides no further asymptotic benefit.

The general question posed by Olesker-Taylor remains open for static
rules outside the distance-unbiased class: does every fixed
distribution over pairs require $\Om{n(\log n)^2}$ proposed exchanges
on some initial permutation with probability tending to one? Allowing
location-dependent selection also leaves a wide gap between scheduled
and adaptive rules. Sorting networks use $\Oh{n\log n}$ predetermined
compare-exchanges~\cite{AjtaiKomlosSzemeredi1983}, whereas unrestricted
rules with complete state information require only $\Oh{n}$ exchanges
when pair selection is uncharged. Intermediate restrictions on location selection and state information
may help explain this gap and yield a more nuanced hierarchy of classes
between these extremes.

\section*{Use of AI tools}
ChatGPT (Astra~6 and Sol~6.1, OpenAI) was used throughout the research process, including literature review, assessment of proofs, and drafting and revising the manuscript. The mathematical arguments were developed through iterative exchanges between the authors and these models. The authors directed the research, evaluated and refined the arguments, and made all final scientific and editorial decisions. The authors checked and revised the AI-assisted content retained in the paper and take full responsibility for its accuracy, originality, and integrity.

\section*{Acknowledgments}
We acknowledge funding by the European Union (ERC, ``dynaBBO'', grant
no.~101125586). Views and opinions expressed are, however, those of the
authors only and do not necessarily reflect those of the European Union or
the European Research Council Executive Agency. Neither the European Union
nor the granting authority can be held responsible for them. This research
was also funded by the Swiss National Science Foundation, grant
no.~200021-232060.

\bibliographystyle{splncs04}
\bibliography{refs}

\end{document}